\documentclass[journal]{IEEEtran}
\IEEEoverridecommandlockouts
\usepackage{cite}
\usepackage{amsmath,amssymb,amsfonts}
\usepackage{algorithmic}
\usepackage{graphicx}
\usepackage{textcomp}
\usepackage{xcolor}
\usepackage{bm}
\usepackage{ulem}
\def\BibTeX{{\rm B\kern-.05em{\sc i\kern-.025em b}\kern-.08em
		T\kern-.1667em\lower.7ex\hbox{E}\kern-.125emX}}

\begin{document}
	
	\title{General Reference Frame Identification and Transformation in Unbalanced Power Systems}
    \author{
    \IEEEauthorblockN{Francisco G. Montoya and }
    \and
    \IEEEauthorblockN{Santiago Sanchez-Acevedo}
\thanks{F. G. Montoya is with the Department
of Electrical Engineering, University of Almeria, Almeria,
Spain (e-mail: pagilm@ual.es).}
\thanks{S. Sanchez-Acevedo is with Department of Energy Systems, SINTEF Energi AS, Trondheim, Norway (e-mail: santiago.sanchez@sintef.no).}
\thanks{Manuscript received June yy,  2025; revised DSDS yy, 2025.}
}

	\maketitle

    \begin{abstract}
        Coordinate transformations provide dimensional reduction benefits across power system analysis, electric machine modeling, and power electronic converter control. This paper introduces a novel transformation based on Geometric Algebra that directly identifies the plane containing unbalanced quantity loci through bivector analysis. The method provides a direct transformation valid for any degree of unbalance in $n$-phase, $(n+1)$-wire sinusoidal systems, requiring only two voltage or current measurements at different time instants. Through pure geometric reasoning, we demonstrate that our approach generalizes existing techniques while extending naturally to multi-dimensional systems. Experimental validation using real-time digital simulation and physical laboratory testing confirms the method's effectiveness under realistic conditions. Power electronics converter control implementation demonstrates significant practical advantages, eliminating zero component oscillations present in Clarke transformation under unbalanced conditions and enabling more effective control architectures. The combination of computational efficiency, robustness, and practical applicability represents a significant advancement for power system control applications.
    \end{abstract}

	
\section{Introduction}

    Power system analysis, modeling, and control rely fundamentally on appropriate coordinate transformations. The Clarke transformation exemplifies this principle by converting three-phase \(abc\) quantities into a different stationary \(\alpha\beta0\) frame \cite{orourke2019}. Extending this idea, \cite{willems2007generalized} generalizes the approach to \(n\)-phase systems. For balanced systems, this transformation achieves dimensional reduction because the locus of the space vector signal inherently lies on the $\alpha\beta$ plane (plane with normal vector [1,1,1]), rendering the remaining coordinates zero.

    The effectiveness of the Clarke transformation diminishes significantly when applied to unbalanced systems. Under unbalanced conditions, the spatial trajectory of quantities deviates from the \(\alpha\beta\) plane and instead occupies a tilted plane whose orientation depends on the unbalance characteristics. Consequently, coordinates different from $\alpha$ and $\beta$ in the general case, become non-zero, preventing the desired dimensional reduction.

    To overcome these limitations, researchers have developed various alternative transformation approaches. The $mno$ transformation by Montanari and Gole \cite{montanari2017} employs an adaptive reference frame that tracks the instantaneous voltage vector, though it becomes undefined when phase voltages reach zero magnitude and is limited to three-phase systems. Tan and Sun \cite{tan2017} introduced a non-orthogonal coordinate system for handling unbalanced conditions, but their method fails under specific scenarios: when phase magnitudes become zero or when phase differences equal 0 or \(\pi\) radians, and is also restricted to three-phase applications. The Reduced Reference Frame (RRF) developed by Casado-Machado et al. \cite{casado2020} aligns the coordinate system with the principal axes of the voltage trajectory's elliptical locus and can be extended to $n$-phase systems. While the transformation shows promise, its dependence on prior amplitude and phase parameter estimation, variable computational burden due to trajectory classification, and validation limited to steady-state scenarios suggest further investigation is needed for dynamic real-time applications. The vector locus transformation presented in \cite{vectorlocus} achieves both null 0-coordinate and constant-valued signals, but remains restricted to three-phase applications and imposes angular constraints on base vector selection. Additionally, \cite{vectorlocus} requires waiting $\pi/2$ radians  i.e. one-quarter electrical cycle to determine the second base vector. Furthermore, this method also relies on preliminary amplitude and phase parameter estimation, introducing similar computational delays and error propagation issues.

    Despite these advances, most existing methods suffer from various constraints: singular behavior under certain operating conditions, dependence on full-period signal measurements, or inability to generalize beyond three-phase systems (with RRF being a notable exception that extends to $n$-phase systems).

    The contribution of this paper lies in the development of a novel transformation methodology based on the foundational principles of Geometric Algebra (GA). We refer to it as \textit{Montoya Transform}. Our approach leverages the bivector representation of the plane containing the unbalanced quantity locus in $n$-dimensional space. We derive a geometric rotor that transforms this arbitrary plane to a canonical reference plane (the $ps$-plane), yielding a transformation valid across all unbalance conditions for $n$-phase, $(n+1)$-wire systems while requiring only two voltage or current samples at a short distinct time instances. The proposed framework not only generalizes seamlessly to multi-dimensional systems but also achieves significant computational savings by requiring fewer measurements—eliminating singularities inherent to conventional approaches.

    The structure of the paper proceeds as follows. Section II reviews essential geometric algebra concepts; Section III presents our proposed transformation methodology; Section IV analyzes the advantages relative to existing approaches; Section V demonstrates the method through numerical validation; and Section VI provides concluding remarks.
	
\section{Geometric Algebra Preliminaries}
	This section provides a brief introduction to the key principles of Geometric Algebra essential to the methodology presented. For a comprehensive treatment of Geometric Algebra, the reader is referred to \cite{hestenes2012clifford,doran2003,macdonald2010linear}.
	
	\subsection{Bivectors and Plane Representation}
    \begin{figure}
        \centering
            \includegraphics[width=\columnwidth]{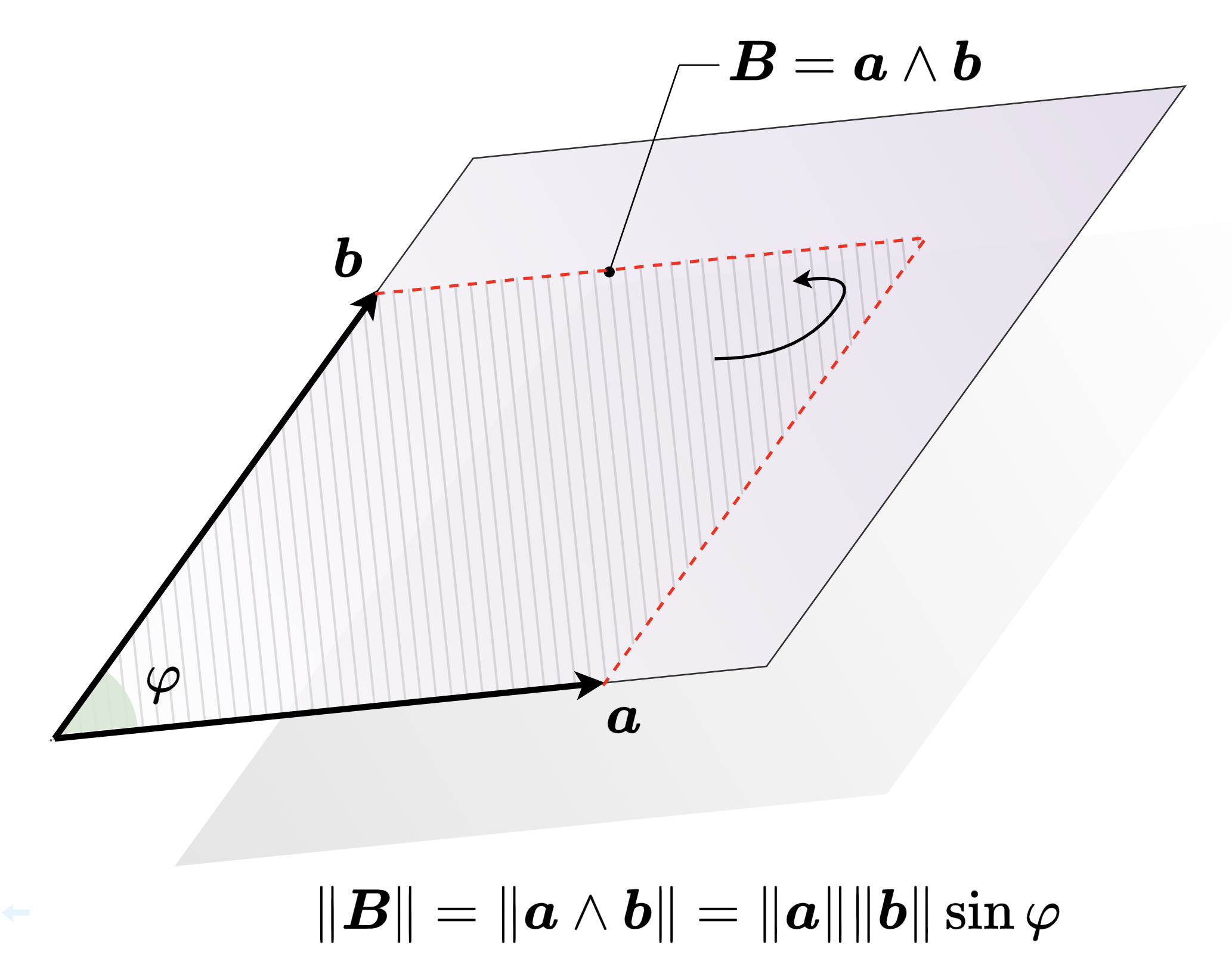}
        \caption{Representation of a plane by a bivector. The wedge product of $\bm{a}$ and $\bm{b}$ is a bivector $\bm{B}$.}
        \label{fig:bivector}
    \end{figure}
	In GA, bivectors are fundamental geometric entities that can represent oriented planes. A (simple) bivector $\bm{B} = \bm{a} \wedge \bm{b}$ is formed through the outer product ($\wedge$) of two vectors, encoding both the orientation of the plane spanned by vectors $\bm{a}$ and $\bm{b}$ and its area (see Fig. \ref{fig:bivector}). The magnitude of the bivector equals the area of the parallelogram formed by the vectors $\bm{a}$ and $\bm{b}$

    \begin{equation}
    	\|\bm{B}\| = B = \|\bm{a}\| \|\bm{b}\|\sin \varphi = ab \sin \varphi
    \end{equation}
    
    \noindent with $\varphi$ the angle between $\bm{a}$ and $\bm{b}$. Bivectors can be normalized to become unit bivectors $\hat{\bm{B}} =\bm{B}/B$. Note that bold letters denote vectors (lowercase) and multivectors (capital letters), while non-bold letters represent their magnitudes or scalars. 
 
    Simple bivectors naturally represent planes in any dimension, making them ideal for our analysis of sinusoidal multi-phase systems where voltage and current loci form planar curves in $n$-dimensional space. For example, the standard basis bivectors in 3D space are $\bm{\sigma}_{12} = \bm{\sigma}_1 \wedge \bm{\sigma}_2$, $\bm{\sigma}_{23} = \bm{\sigma}_2 \wedge \bm{\sigma}_3$, and $\bm{\sigma}_{31} = \bm{\sigma}_3 \wedge \bm{\sigma}_1=-\bm{\sigma}_{13}$.
	
	\subsection{Multivectors and the Geometric Product}
	GA is a mathematical framework that provides a unified representation of geometric entities through multivectors. A multivector can contain elements of different geometric dimensions (scalars, vectors, bivectors, etc.) in a single algebraic structure \cite{doran2003}.
	
	The geometric product is the fundamental operation in GA. For bivectors, which are central to our approach, the geometric product between two bivectors $\bm{A}$ and $\bm{B}$ can be decomposed into components of different grades:
	
	\begin{equation}
		\begin{aligned}
			\bm{AB} &= \langle\bm{AB}\rangle_0 + \langle\bm{AB}\rangle_2 + \langle\bm{AB}\rangle_4 \\
			&= \bm{A} \rfloor \bm{B} + \bm{A}\times \bm{B} + \bm{A} \wedge \bm{B}
		\end{aligned}
	\end{equation}	
	\noindent where $\langle\bm{AB}\rangle_k$ denotes the grade-$k$ part of the product. The term $\bm{A} \rfloor \bm{B}=\langle\bm{AB}\rangle_0$ is the left contraction product between bivectors (very similar to the classical inner/dot product between vectors, but generalized for $k$-vectors, see \cite{dorst2002inner}),  $\bm{A}\times \bm{B}$ is the commutator product (similar to Lie bracket operation and  different from traditional vector product) resulting in a bivector part, and $\bm{A} \wedge \bm{B}=\langle\bm{AB}\rangle_4$ is the outer product resulting in a grade-4 element (when it exists in the space).
	
	When bivectors $\bm{A}$ and $\bm{B}$ intersect (as is always the case in $\mathbb{R}^3$), we have $\bm{A} \wedge \bm{B} = 0$, and the geometric product simplifies to:	
	\begin{equation}
		\bm{AB} = \bm{A} \rfloor \bm{B} +  \bm{A}\times \bm{B}
	\end{equation}	
	\noindent with	
	\begin{equation*}
		\begin{aligned}
			\bm{A} \rfloor \bm{B}  = \frac{\bm{A} \bm{B} + \bm{B}  \bm{A}  }{2}\qquad 
			\bm{A} \times \bm{B}  = \frac{\bm{A} \bm{B} - \bm{B}  \bm{A}  }{2}
		\end{aligned}
	\end{equation*}
	
	The angle between two planes represented by bivectors $\bm{A}$ and $\bm{B}$ can be computed using \cite{hitzer2012}:
	
	\begin{equation}
		\cos \theta = \frac{\bm{A} \rfloor {\bm{B}^{\dagger}}}{AB}= \frac{\langle \bm{A} {\bm{B}^{\dagger}} \rangle_0}{AB} 
		\label{eq:cos_angle}
	\end{equation}
	
	where ${\bm{B}}^{\dagger}$ is the reverse operation on $\bm{B}$, and $\langle \cdot \rangle_0$ denotes the scalar part. This is a general formula and holds for $n$-dimensional spaces.
	
	\subsection{Rotors and Rotations}
	In GA, rotations are elegantly represented using geometric rotors, which are specific multivectors that satisfy $\bm{R}\bm{R}^{\dagger} = 1$. 
	
	When two planes  $\bm{A}$ and $\bm{B}$ (with unit bivectors $\hat{\bm{A}}$ and $\hat{\bm{B}}$) intersect, the rotor that rotates from $\bm{B}$ to plane $\bm{A}$ can be expressed as:
	
	\begin{equation}
		\bm{R} = \frac{1+\hat{\bm{A}}\hat{\bm{B}}^{\dagger}}{\|1+\hat{\bm{A}}\hat{\bm{B}}^{\dagger}\|}= \exp\left(\frac{\theta\hat{\bm{L}}}{2}\right)
		\label{eq:rotor}
	\end{equation}
	
	where $\hat{\bm{L}}$ is a unit bivector representing the plane of rotation, and $\theta$ is the rotation angle. The plane of rotation $\hat{\bm{L}}$ is perpendicular to both $\bm{A}$ and $\bm{B}$ and obviously to their  intersection line.  It can be computed from the bivector part (the commutator product) of their geometric product using
	
	\begin{equation}
		\hat{\bm{L}} = \frac{\bm{A}\times \bm{B}^{\dagger}}{\| \bm{A}\times \bm{B}^{\dagger}\|}
		\label{eq:plane_rotation}
	\end{equation}
	
	A general multivector $\bm{X}$ is rotated by a rotor $\bm{R}$ through the sandwich product:
	
	\begin{equation}
		\bm{X'} = \bm{R}\bm{X}\bm{R}^{\dagger}
		\label{eq:transform}
	\end{equation}
	
	This formula applies universally to vectors, bivectors, and any multivector, making it particularly powerful for our transformations of multi-phase quantities.

	\section{Proposed Transformation Method}
	\label{sec:method}
	In GA, the standard orthonormal basis of $n$-dimensional Euclidean space  $\{ \bm{\sigma}_1,\bm{\sigma}_2,\ldots,\bm{\sigma}_n \}$ is used to accommodate the typical $n$ phases values of a power system (voltages, currents, etc). Thus, $n$-phase electrical quantities can be represented as:
	
	\begin{equation}
		\bm{v}(t) = \sum v_k(t)\bm{\sigma}_k
        \label{eq:n-dimensional_vector}
	\end{equation}
	
	This representation allows us to utilize GA operations, particularly the outer product and rotors, to develop our transformation method.

	\subsection{Unbalanced Sinusoidal Three-Phase System}
	We start with an unbalanced sinusoidal three-phase system described by the following voltages:
	
	\begin{equation}
		\begin{aligned}
			v_a(t) &= V_a \cos(\omega t + \varphi_a) \\
			v_b(t) &= V_b \cos(\omega t + \varphi_b) \\
			v_c(t) &= V_c \cos(\omega t + \varphi_c)
		\end{aligned}
        \label{eq:trhee_phase_signal}
	\end{equation}

	where $V_a$, $V_b$, and $V_c$ are the amplitudes, and $\varphi_a$, $\varphi_b$, and $\varphi_c$ are the phase angles. From now on, we focus on voltages, but the same rationale can be applied to currents. 
	
	
	In balanced systems, where $V_a = V_b = V_c$ and $\varphi_a - \varphi_b = \varphi_b - \varphi_c = \varphi_c - \varphi_a = 2\pi/3$, the curve described by $\bm{v}(t)$ is a circle that lies in a plane with normal vector $\left[1,1,1\right]$. This is precisely the plane targeted by the Clarke transformation. However, in unbalanced systems, the locus remains a planar curve (an ellipse), but the plane is generally different.
	
\subsection{Identifying the Plane of the Voltage Locus}

     Combining Eqs. \eqref{eq:n-dimensional_vector} and \eqref{eq:trhee_phase_signal}, a three phase vector signal can be expressed as  (see \cite{eid2022geometric})

        \begin{equation}
        \boldsymbol{v}(t)= \cos (\omega t) \boldsymbol{p}-\sin (\omega t) \boldsymbol{s}
        \end{equation}
        \noindent with
        \begin{align}
        \boldsymbol{p} & =\sum_{k=1}^3 V_k \cos \varphi_k\boldsymbol{\sigma}_k \\
        \boldsymbol{s} & =\sum_{k=1}^3 V_k \sin \varphi_k \boldsymbol{\sigma}_k
        \end{align}
    Our approach directly identifies the plane containing the voltage locus using geometric algebra. The key insight is that any two non-collinear voltage vectors are sufficient to determine this plane. We use $\bm{p}$ and $\bm{s}$ for this task.

    The bivector $\bm{B}$ representing the plane ($ps$-plane) containing the voltage locus is computed as:
	
    	\begin{equation}
    		\begin{aligned}
    			\bm{B} &= \bm{p} \wedge \bm{s} 
    			= B_{12}\bm{\sigma}_{12}  +B_{13}\bm{\sigma}_{13} + B_{23}\bm{\sigma}_{23}
    		\end{aligned}
    		\label{eq_wedgeB}
    	\end{equation}
    \noindent where $\bm{\sigma}_{ij} = \bm{\sigma}_i \wedge \bm{\sigma}_j$ are the basis bivectors and the components $B_{ij}$ are 
	\begin{equation}
		\begin{aligned}
			B_{12} &= V_a V_b \sin(\varphi_a - \varphi_b) \\
			B_{13} & = V_a V_c \sin(\varphi_a - \varphi_c)\\
			B_{23} & =  V_b V_c \sin(\varphi_b - \varphi_c)
		\end{aligned}	
		\label{eq:bivector_components}
	\end{equation}
    For practical purposes, it suffices to take any two non-collinear voltage vectors (let's say $\bm{v}_1$ and $\bm{v}_2$ with $\bm{v}_1 \wedge \bm{v}_2 \neq 0$) measured at different time instants $t_1$ and $t_2$ (with $t_2 > t_1$ and $t_2-t_1\leq T$) to determine $\bm{B}$. It is important to remark that the instant $t_2$ can be arbitrary, which avoids the quarter-of-a-period restriction described in \cite{vectorlocus}. However, the choice $t_2 = t_1+T/2$ leads to a singularity in the problem (not in the method), as the voltage vectors become collinear, thus contained in an infinite number of planes.

	For example, taking two measurements of the voltage vector at times $t_1$ and $t_2$
	\begin{equation}
		\begin{aligned}
			\bm{v}_1 &= v_{a1}\bm{\sigma}_1 + v_{b1}\bm{\sigma}_2 + v_{c1}\bm{\sigma}_3 \\
			\bm{v}_2 &= v_{a2}\bm{\sigma}_1 + v_{b2}\bm{\sigma}_2 + v_{c2}\bm{\sigma}_3
		\end{aligned}
	\end{equation}

	\noindent the bivector $\bm{B}$ representing the plane containing the voltage locus can be computed as
	
	\begin{equation}
		\begin{aligned}
			\bm{B} &= \bm{v}_1 \wedge \bm{v}_2 
			= (v_{a1}v_{b2} - v_{b1}v_{a2})\bm{\sigma}_{12} \\
			&+  (v_{a1}v_{c2} -v_{c1}v_{a2}) \bm{\sigma}_{13}  
			 + (v_{b1}v_{c2} - v_{c1}v_{b2})\bm{\sigma}_{23}
		\end{aligned}
	\end{equation}
	
	Note that if $\bm{v}_1$ and $\bm{v}_2$ are always collinear (irrespective of the chosen instant of time), then $\bm{B}=0$, and thus, we recognize the zero sequence component. 
	For balanced systems, $\bm{B}$ is proportional to $\bm{\sigma}_{12} + \bm{\sigma}_{13} + \bm{\sigma}_{23}$, which represents a plane consistent with the Clarke transformation.

	\subsection{Computing the Rotor for Plane Alignment}
	Our objective is to find a rotor that rotates the $ps$-plane containing the voltage locus (represented by ${\bm{B}}$) to align with one of the canonical planes---$\bm{\sigma}_{12}$ ($xy$-plane), $\bm{\sigma}_{13}$ ($xz$-plane) or $\bm{\sigma}_{23}$ ($yz$-plane)---thus achieving the dimensionality reduction from 3 to 2 coordinates. For convenience, we arbitrarily choose the bivector $\bm{\sigma}_{12}$, which means it effectively discards the $z$-coordinate in the transformed system. 
	
	The rotor that performs this rotation can be computed using Eqs. \eqref{eq:cos_angle} to \eqref{eq:plane_rotation},  where $\bm{A} = \hat{\bm{A}}=\bm{\sigma}_{12} $, and with parameters

	\begin{equation}
		\cos \theta = \frac{B_{12}}{\sqrt{B_{12}^2+B_{13}^2+B_{23}^2}}
		\label{eq:angle}
	\end{equation}
	
	\begin{equation}
		\hat{\bm{L}} = \frac{-B_{23}\bm{\sigma}_{13} + B_{13}\bm{\sigma}_{23}}{\sqrt{B_{13}^2+B_{23}^2}}
		\label{eq:L_plane}
	\end{equation}
	
		
		\subsection{Complete Transformation}
		With the rotor $\bm{R}$ computed, the complete transformation of the three-phase vector $\bm{v}$ consists of:
		
		\begin{equation}
			\bm{v'} = \bm{R}\bm{v}\bm{R^{\dagger}}
		\end{equation}
		
		The transformed vector $\bm{v'}$ has its third component ($z$-component) equal to zero, as it now lies entirely in the $\bm{\sigma}_{12}$-plane. In the balanced case, this transformation is equivalent to the SKR transformation \cite{Eid2022}. In the unbalanced case, it provides the optimal transformation that maps the voltage locus to a two-dimensional one.

	\subsection{Extension to $n$-Dimensional Systems}
	A significant advantage of our GA approach is its natural extension to $n$-dimensional systems. In an $n$-phase system, the voltage vector is represented as in \eqref{eq:n-dimensional_vector}.	
	The locus of this vector generally lies in a two-dimensional plane embedded in the $n$-dimensional space. By taking two measurements of the voltage vector, we can compute the bivector representing this plane:
	
	\begin{equation}
		{\bm{B}} = \bm{v}_1 \wedge \bm{v}_2
	\end{equation}
	
	This bivector can be expressed in terms of the basis bivectors $\bm{\sigma}_{ij}$:
	
	\begin{equation}
		{\bm{B}} = \sum_{i<j}^n {B}_{ij}\bm{\sigma}_{ij}
	\end{equation}
	
	When extending to $n$ dimensions where $n > 3$, we face an important geometric difference: two planes may not intersect in a line, unlike in three dimensions where any two planes always do. This means that the direct application of equation \eqref{eq:rotor} is not always possible since it assumes intersection between planes. To overcome this challenge, we employ a two-step rotation process. 
	
	\subsubsection{Step 1. Rotating $\bm{v}_1$ to $\bm{\sigma}_1$}
	We first normalize $\bm{v}_1$ to obtain a unit vector:
	
	\begin{equation}
		\hat{\bm{v}}_1 = \frac{\bm{v}_1}{||\bm{v}_1||}
	\end{equation}
	
	The rotor $\bm{R}_1$ that rotates $\hat{\bm{v}}_1$ to $\bm{\sigma}_1$ has the same structure as for bivectors, described in \eqref{eq:rotor}:
	
	\begin{equation}
		\bm{R}_1 = \frac{1 + \bm{\sigma}_1\hat{\bm{v}}_1}{\|1 + \bm{\sigma}_1\hat{\bm{v}}_1\|}
        \label{eq:rotorR1}
	\end{equation}
	
	This formula works in any dimension and creates a minimal rotation that aligns $\hat{\bm{v}}_1$ with $\bm{\sigma}_1$.
	
	\subsubsection{Step 2. Rotate plane $\bm{B}$ and align the transformed plane ${\bm{B}_{\times}}$ with $\bm{\sigma}_{12}$}
	After obtaining $\bm{R}_1$, we transform the original unit bivector $\hat{\bm{B}}$ to the new plane :
	
	\begin{equation}
		\hat{\bm{B}}_{\times} = \bm{R}_1 \hat{\bm{B}} \bm{R}_1^{\dagger}
	\end{equation}
	
	By construction, $\hat{\bm{\bm{B}}}_{\times}$ contains $\bm{\sigma}_1$ (since $\bm{R}_1$ rotates $\bm{v}_1$ to $\bm{\sigma}_1$), which means $\hat{\bm{B}}_{\times}$ must intersect with the $\bm{\sigma}_{12}$ plane along the $\bm{\sigma}_1$ direction. 
	Since $\hat{\bm{B}}_{\times}$ and $\bm{\sigma}_{12}$ now intersect, we can apply again equation \eqref{eq:rotor} to find a second rotor $\bm{R}_2$:
	
	\begin{equation}
		\bm{R}_2 = \frac{1 + \bm{\sigma}_{12}\hat{\bm{B}}_{\times}^{\dagger}}{||1 + \bm{\sigma}_{12}\hat{\bm{B}}_{\times}^{\dagger}||}
        \label{eq:rotorR2}
	\end{equation}
	
	\subsubsection{Step 3. Complete Transformation}
	The complete rotor that aligns the original plane $\bm{B}$ with the target plane $\bm{\sigma}_{12}$ is the product of the two rotors:
	
	\begin{equation}
		\bm{R} = \bm{R}_2 \bm{R}_1
	\end{equation}
	
	This rotor can then be applied to transform any vector $\bm{v}$ in the original $n$-dimensional space:
	
	\begin{equation}
		\bm{v}' = \bm{R} \bm{v} \bm{R}^{\dagger}
        \label{eq:vector_rotated}
	\end{equation}
	
	The transformed vector $\bm{v}'$ will have all components beyond the first two effectively reduced to zero, achieving a dimensionality reduction from $n$ to 2, regardless of the original number of phases.
	This straightforward extension to $n$ dimensions is a direct consequence of the geometric algebra framework, which provides consistent operations across spaces of any dimension. None of the previously proposed methods for unbalanced three-phase systems can be so naturally extended to higher dimensions.
   
\section{Advantages over Existing Methods}
	Our geometric algebra approach offers several key advantages over existing methods – some of which extend beyond the benefits typically highlighted in prior works.
		
	\textbf{Robustness}: The method is valid for any degree of unbalance, introducing no artificial singularities beyond those inherent to the physical system due to signal colinearities. Unlike methods based on other analytical formulations like the $mno$ transformation \cite{montanari2017} or the Tan-Sun coordinate transformation \cite{tan2017}, our approach does not fail when any phase has zero voltage magnitude or when the phase difference between any two phases is 0 or $\pi$ radians.
			
	\textbf{Simplicity}: Only two voltage/current measurements are needed to compute the transformation. This is in contrast to methods like \cite{casado2020} and \cite{vectorlocus}, which require multiple steps and specific measurement timings or parameter estimation. 
			
	 \textbf{Geometric insight}: The bivector representation provides clear geometric interpretation of the unbalance. The magnitude of the angle between the bivector $\bm{B}$ and the $\bm{\sigma}_{12}$-plane directly quantifies the degree of unbalance, providing valuable diagnostic information.
			
    \textbf{Computational efficiency}: The transformation can be computed with minimal computational resources. The core operations (outer product, normalization, and rotor calculation) are straightforward to implement and can be executed in real-time on standard microcontrollers.
			
    \textbf{Generality}: The approach generalizes the classical transformations while providing a more robust framework. In the balanced case, our transformation reduces to the classical Clarke transformation, ensuring backward compatibility with existing control systems.
			
	\textbf{Extensibility to $n$ dimensions}: Unlike the majority of the previously proposed methods that are inherently limited to three-phase systems, our geometric algebra approach extends naturally to $n$-dimensional spaces. The core concepts of bivectors, outer products, and rotors are all well-defined in arbitrary dimensions. This allows our method to be directly applicable to multi-phase systems (e.g., five-phase, six-phase, or higher) without requiring fundamental reformulation.


    \section{Validation and Applications}
    To demonstrate the effectiveness of the proposed method, we present comprehensive validation through analytical examples, control applications, and experimental testing. We begin with a detailed three-phase unbalanced system analysis to illustrate the core methodology and geometric transformations. Next, we demonstrate the practical application of the method in power electronics converter control, showing how the transformation integrates into existing control architectures. We then validate the approach through experimental laboratory tests using an OPAL-RT real-time digital simulator platform in a realistic power system setup with fault conditions. Finally, we showcase the method's extensibility by applying it to a six-phase synthetic system, demonstrating the natural generalization to higher-dimensional spaces. These validations collectively confirm the method's ability to correctly identify the voltage locus plane and perform dimensional reduction under various unbalance conditions across different system configurations.

    \subsection{Three-Phase System}
    
	To provide initial validation of our approach, we include a case with an unbalanced three-phase voltage system with the following values:
		\begin{align*}
		v_a(t) &= 1.70 \cos(\omega t ) \\ 
			v_b(t) &= 0.70 \cos(\omega t - 2.1) \\
			v_c(t) &= 1.40 \cos(\omega t - 2.2)
		\end{align*}
		\noindent where $\omega = 2\pi f$ with $f = 50$ Hz. This system exhibits significant imbalance in both amplitudes and phase angles.
        
		\subsection{Plane Identification}
		Following the process outlined in Section~\ref{sec:method}, we first identify the $ps$-plane containing the voltage locus. We select two voltage vectors at times $t_1 = 0$ and $t_2 = T/4$ (5ms):
		
		\begin{align*}
			\bm{v}_1 & = 1.70\bm{\sigma}_1 - 0.35\bm{\sigma}_2 - 0.82\bm{\sigma}_3 \\
			\bm{v}_2 & = 0.00\bm{\sigma}_1 + 0.60\bm{\sigma}_2 -1.13\bm{\sigma}_3
		\end{align*}
		
        These vectors have the following properties

        \begin{equation*}
            \|\bm{v}_1\|=1.92, \qquad  \|\bm{v}_2\|=1.28, \qquad  \varphi = 44.23^{\circ}
        \end{equation*}
		
		\noindent where $\varphi$ is the angle between $\bm{v}_1$ and $\bm{v}_2$. The  bivector $\bm{B}$ (representing the $ps$-plane) is calculated using Equation~\eqref{eq_wedgeB}:
		\begin{align*}
			{\bm{B}} = \bm{v}_1 \wedge \bm{v}_2 = 1.027 \bm{\sigma}_{12} -1.924\bm{\sigma}_{13} + 0.897 \bm{\sigma}_{23}
		\end{align*}

			\begin{figure}[]
			\centering
			\includegraphics[width=\columnwidth]{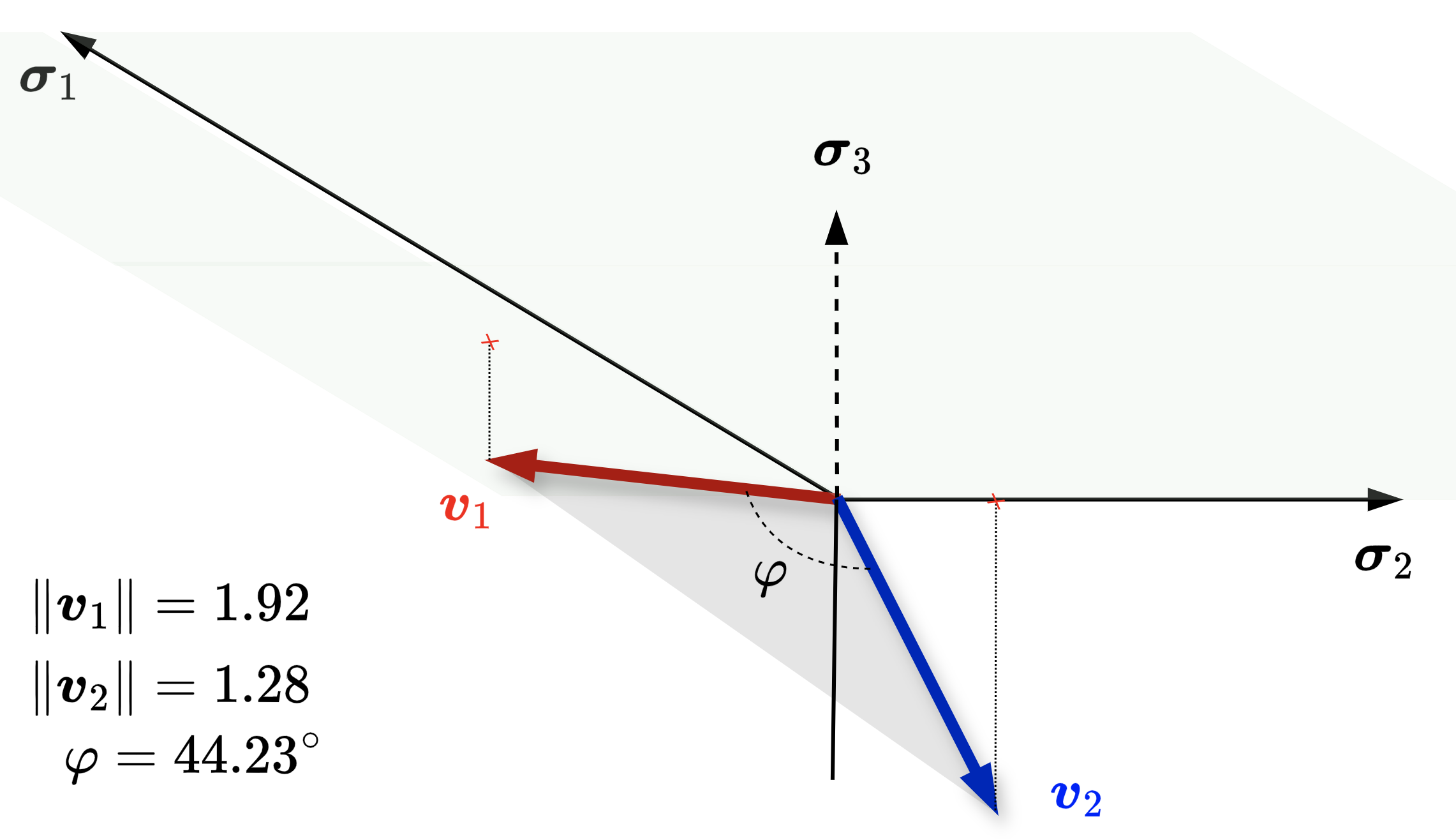}
			\caption{Vector representation of voltage samples at $t_1$ and $t_2$.}
			\label{fig:3dnumerical}
			\end{figure}
		
		The components can also be calculated directly using Equation~(\ref{eq:bivector_components}), yielding the same result. The presence of all three components confirms the non-alignment of the voltage plane with any of the coordinate planes.
		
		\subsubsection{Rotor Calculation}
		The angle between the voltage plane and the target $\bm{\sigma}_{12}$-plane is determined using Equation~(\ref{eq:angle}):
		\begin{align*}
			\cos\theta = \frac{B_{12}}{B} = 0.435
		\end{align*}
		
		yielding $\theta = 64.18^{\circ}$. 
		
		The rotation plane bivector $\hat{\bm{L}}$ is calculated according to Equation~(\ref{eq:L_plane}):
		\begin{align*}
			\hat{\bm{L}} = \frac{-B_{23}\bm{\sigma}_{13} + B_{13}\bm{\sigma}_{23}}{\sqrt{B_{13}^2 + B_{23}^2}} = -0.423 \bm{\sigma}_{13} - 0.906 \bm{\sigma}_{23}
		\end{align*}
	
		Finally, the rotor $\bm{R}$ that performs the alignment is computed using Equation~(\ref{eq:rotor}):
		\begin{align*}
			\bm{R} &= \exp\left(\frac{\theta\hat{\bm{L}}}{2}\right) = \cos\frac{\theta}{2} + \sin\frac{\theta}{2}\hat{\bm{L}} \\
			&= 0.847 -0.225 \bm{\sigma}_{13} - 0.481 \bm{\sigma}_{23}
		\end{align*}

		\subsubsection{Transformation Results}
		Applying the rotor to the voltage vectors using the sandwich product  as in Equation~(\ref{eq:transform}), $\bm{v}_1$ and $\bm{v}_2$ are transferred to the $\bm{\sigma}_{12}$-plane. 
		To verify the transformation's accuracy, we examine the transformed bivector:
		\begin{align*}
			\mathbf{B}' = \mathbf{R}\mathbf{B}\mathbf{R}^{\dagger} = 2.36 \bm{\sigma}_{12} + 0.00 \bm{\sigma}_{13} + 0.00 \bm{\sigma}_{23}
		\end{align*}
		
		The resulting bivector has only the $\bm{\sigma}_{12}$ component, which confirms perfect alignment with the target plane. The transformed vectors are:

        \begin{equation*}
            \begin{aligned}
            \bm{v}_1^{\prime} & = 1.92\bm{\sigma}_1 + 0.12\bm{\sigma}_2\\
			\bm{v}_2^{\prime} & = 0.30\bm{\sigma}_1 + 1.24\bm{\sigma}_2
            \end{aligned}
        \end{equation*}

        \noindent with same lengths and angle that $\bm{v}_1$ and $\bm{v}_2$ but now in the $\bm{\sigma}_{12}$ plane.

\subsection{Power electronics converter controller}
    \begin{figure}
        \centering
        \includegraphics[width=\columnwidth]{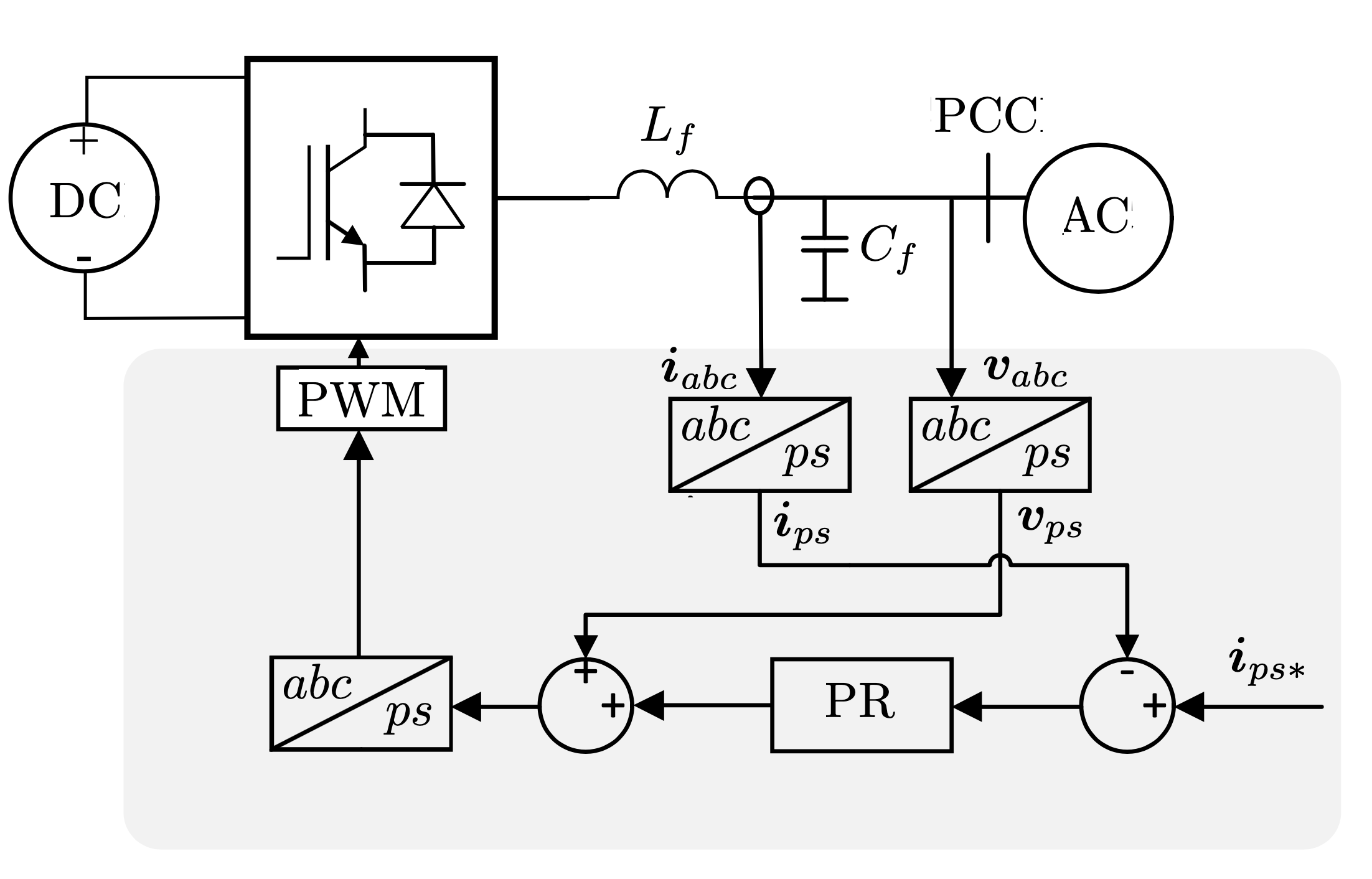}
        \caption{Current control loop for a converter using the GA transform.}
        \label{fig:ControllerPS}
    \end{figure}
    
    This subsection demonstrates the practical application of the proposed transformation to control a voltage source converter (VSC). Fig. \ref{fig:ControllerPS} shows a VSC connected to an equivalent AC grid through an $LC$ filter, where $L_f$ is the filter inductance and $C_f$ is the filter capacitance. The equivalent AC grid has an unbalanced series impedance. The control strategy implements the proposed transformation from $abc$ to $ps$ coordinates as described in Section \ref{sec:method}, thus achieving coordinate reduction.

    The bivector $\bm{B}$ is obtained recursively, where voltage vectors are separated by $\kappa = 8$ samples. Specifically, the voltage vectors are $\bm{v}(t-\kappa T_s)$ and $\bm{v}(t)$, where $T_s$ is the sampling period. The angle between the voltage plane and the target plane $\bm{\sigma}_{12}$ is computed, and the rotor calculation follows the methodology defined in Section \ref{sec:method}.

    The controller follows the same architecture as grid-following converter controllers based on Clarke transform. Current regulation is performed with a proportional plus resonance regulator defined in \eqref{eq:PRcontrol}, where $\rho$ is a damping factor, $\omega$ is the resonance angular frequency, and $k_i$, $k_p$ are the resonance and proportional gains, respectively.
    \begin{equation}
    \label{eq:PRcontrol}
    H(s) = k_p+ \frac{2k_is}{s^2 + 2\rho \omega s + \omega^2}
    \end{equation}
    The reference currents $\bm{i}_{ps*}$ are calculated using the measured voltage $\bm{v}_{ps}$ and a geometric power $\bm{M}$ \cite{montoya2022}. The geometric power $\bm{M}$ is calculated as the geometric product of $\bm{v}_{ps}= v_p\bm{\sigma}_p+ v_s\bm{\sigma}_s$ and $\bm{i}_{ps}=i_p\bm{\sigma}_p+i_s\bm{\sigma}_s$. This yields $\bm{M} = \bm{v}_{ps}\bm{i}_{ps} = p_0+\bm{N}$, where $p_0 = \bm{v}_{ps}\cdot \bm{i}_{ps}$ represents the scalar power and $\bm{N}=\bm{v}_{ps}\wedge \bm{i}_{ps}$ represents the bivector power component.
    
\begin{figure}
        \centering
    \includegraphics[scale=0.6,clip,trim=3cm 8.5cm 3cm 8cm]{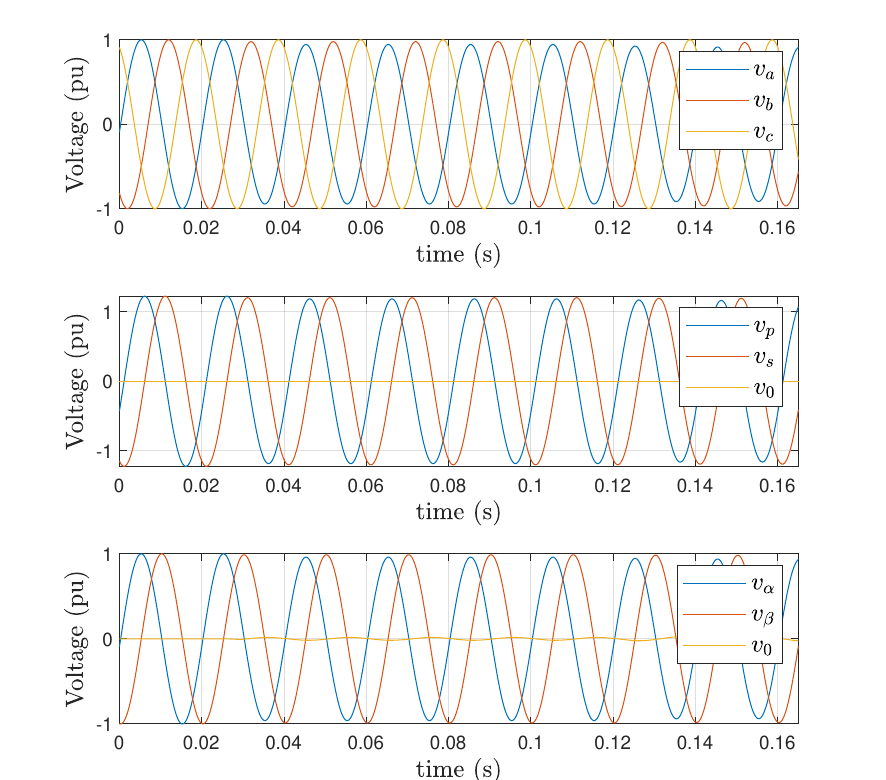}
        \caption{Voltage behavior with abc frame (top), $ps$ frame (middle) and $\alpha\beta$ frame (bottom).}
        \label{fig:voltagecontroller}
    \end{figure}

\begin{figure}
        \centering
    \includegraphics[scale=0.6,clip,trim=3cm 8.5cm 3cm 8cm]{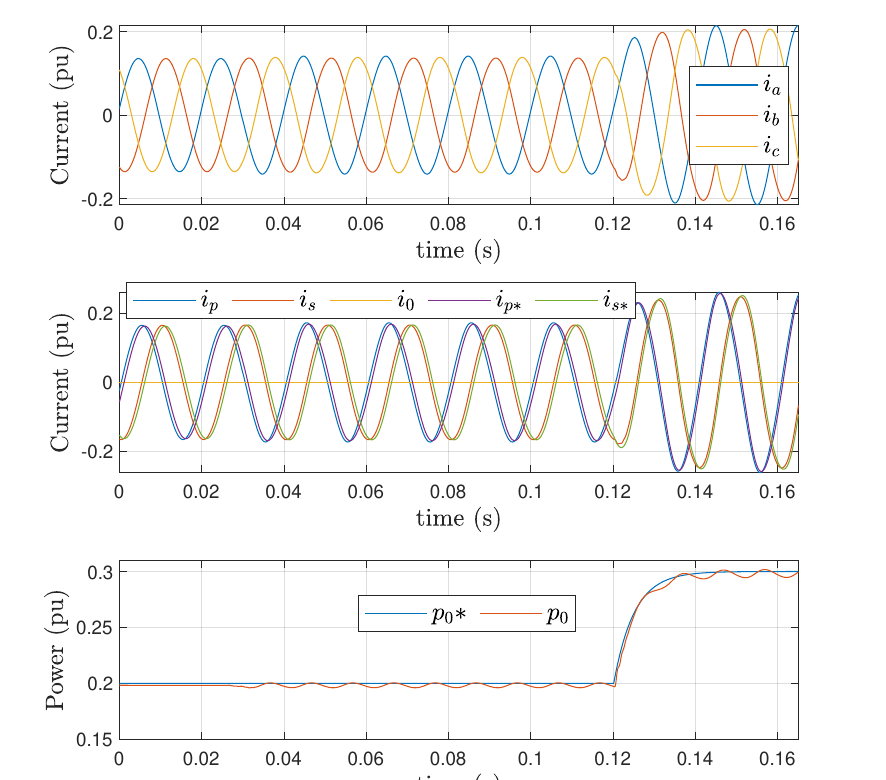}
        \caption{Current and instantaneous power behavior.}
        \label{fig:currentcontroller}
    \end{figure}

\begin{figure}
        \centering
    \includegraphics[scale=0.6,clip,trim=3cm 9.7cm 2cm 10cm]{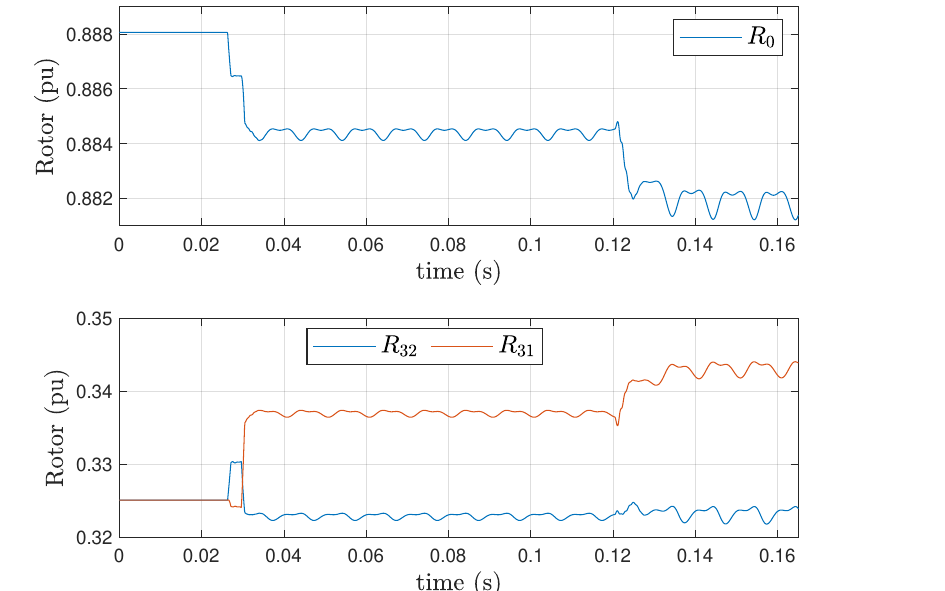}
        \caption{Rotor parameters, scalar (top), grade 2 elements of the rotor (bottom).}
        \label{fig:rotorControlps}
    \end{figure}

    The controller performance was evaluated through simulations conducted in MATLAB Simulink with $T_s = 100\ \mu$s, with results presented in Figs. \ref{fig:voltagecontroller}, \ref{fig:currentcontroller}, and \ref{fig:rotorControlps}. The test scenario involves the introduction of an unbalance in the three-phase voltages at $t = 0.02$ s. At $t = 0.12$ s, there is a step change in the power reference that produces a change in the current injected into the AC grid. This increase in current further increases the voltage unbalance due to the equivalent unbalanced series impedance. These unbalanced conditions produce a non-zero $v_0$ component in the $\alpha\beta$ frame.

    Fig. \ref{fig:voltagecontroller} (middle) shows that under voltage unbalance conditions, the $ps$ transformation does not produce any zero-sequence component. In contrast, the Clarke transformation shows significant oscillations in the zero component $v_0$ starting from 0.02 s (Fig.\ref{fig:voltagecontroller}, bottom).  

    Fig. \ref{fig:currentcontroller} (top) shows the three-phase current waveforms, while the $ps$ currents and their references are shown in Fig. \ref{fig:currentcontroller} (middle). The reference and measured instantaneous scalar power are presented in Fig. \ref{fig:currentcontroller} (bottom). The power reference step is applied at $t = 0.12$ s. The results demonstrate effective power tracking performance with the proposed $ps$ transformation-based controller.

    The components of the rotor used to transform the three-phase voltages and currents are presented in Fig. \ref{fig:rotorControlps}, noting that the component $R_{12}=0$ by design. The activation of the voltage unbalance at 0.02 s produces a distinct change in the remaining components of $\mathbf{R}$. A second change is observed at 0.12 s in response to the power reference change in the controller.

\subsection{Experimental laboratory tests}
    \begin{figure}
	   \centering 
        \includegraphics[clip,trim= 4.7cm 13cm 4cm 0.2cm,scale=0.42]{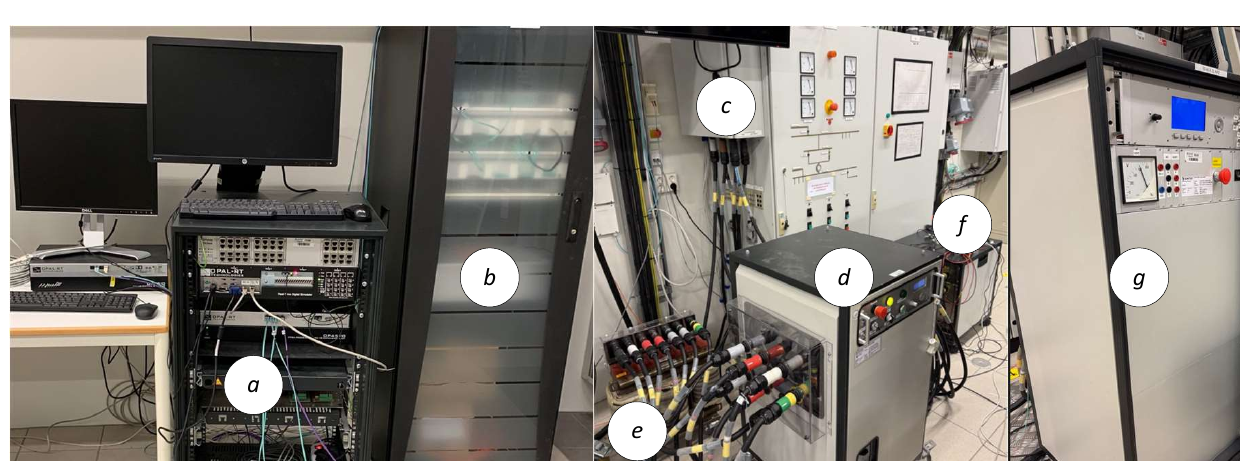}
	    \caption{Laboratory and setup. a) OPAL-RT, b) optic fiber rack, c) main grid, d)Fault emulator device, e) Reactance, f) measurement box and g) voltage source converter.}
	    \label{fig:nsgl}
	\end{figure}
\begin{figure}
        \centering
        \includegraphics[width=1\linewidth]{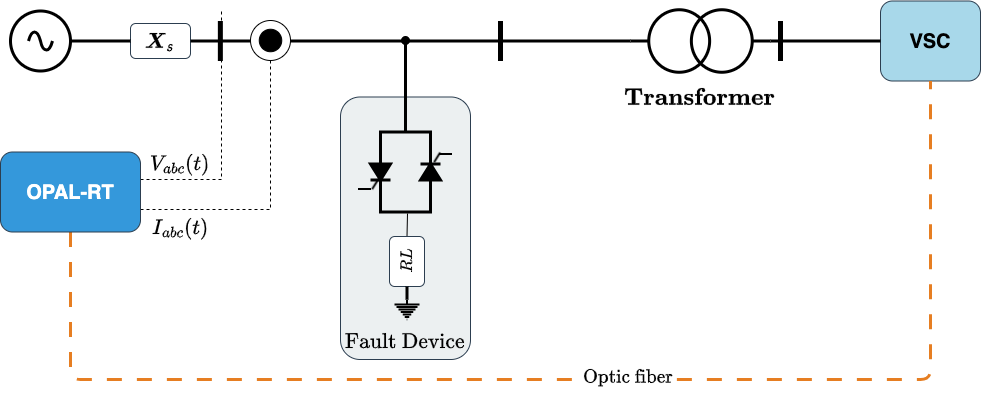}
        \caption{Scheme of the circuit used in the experimental validation.}
        \label{fig:schemaLab}
    \end{figure}

    The aim of this experimental validation is to show the performance of the method and the correct behavior in realistic environments. 
    The laboratory setup described in Fig. \ref{fig:nsgl} has been implemented in the Norwegian Smart Grid laboratory (see \cite{NSGL}). The schematic of the setup is shown in Fig. \ref{fig:schemaLab}. A thyristor-based fault emulator with a reactor has been connected to the main grid to recreate a fault and a 60 kVA power electronics voltage source converter (VSC) has been connected to the system to obtain the power exchange. Hence, this setup follows the characteristics of modern power systems with converter-based renewable energy. The OPAL-RT digital real-time simulator is used for data logging and control of the VSC. In addition, a phase-phase-ground fault is used to demonstrate the computational efficiency of the method.

\begin{figure}
        \centering
            \includegraphics[width=\columnwidth]{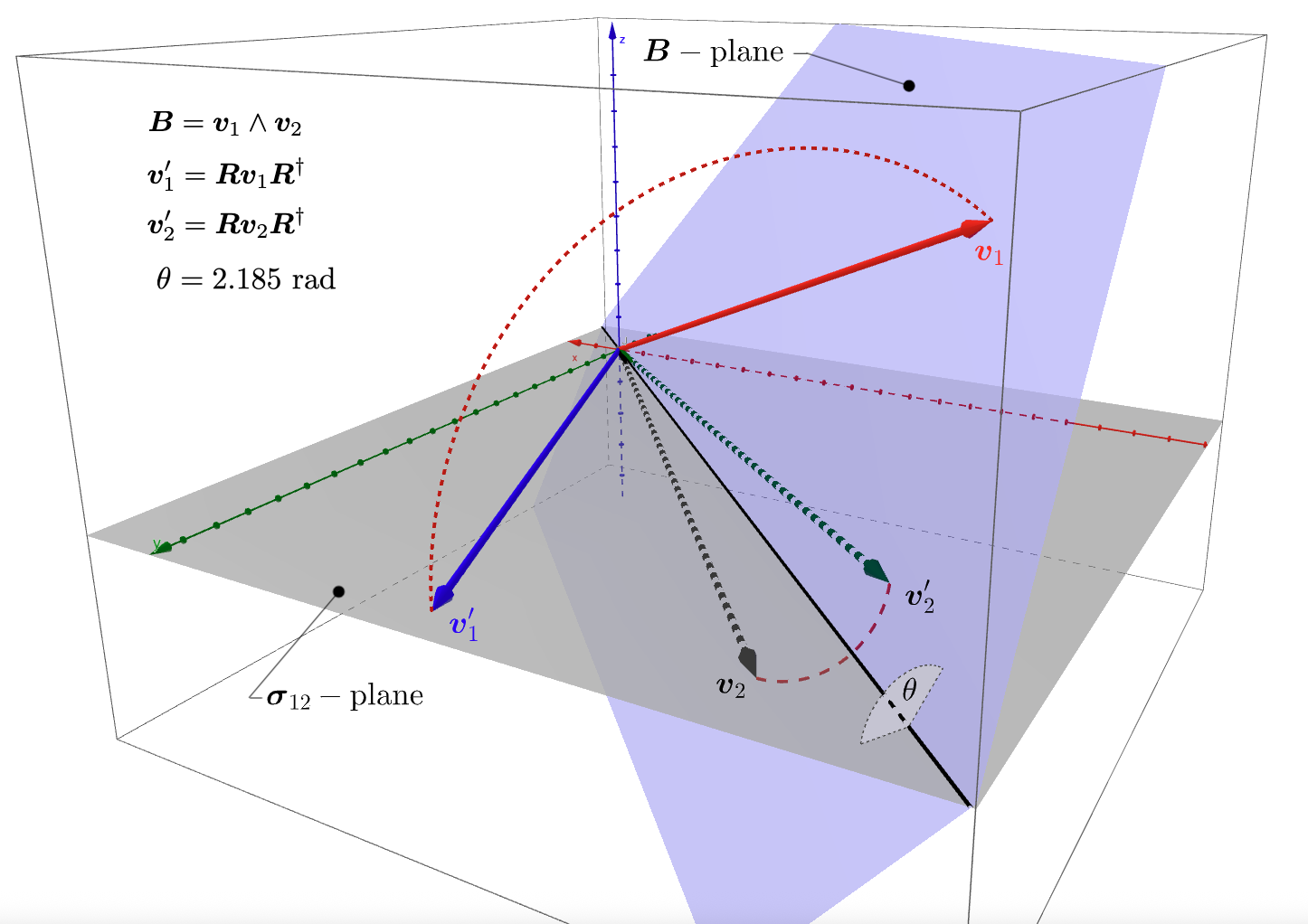}
        \caption{Geometric representation of 2 voltage vectors $\bm{v}_1$ and $\bm{v}_2$ for different instants of time. The computed plane $\bm{B}$ ($ps$-plane) and the rotated vectors $\bm{v}_1^{\prime}$ and $\bm{v}_2^{\prime}$ are also shown. Note that $\bm{v}_1$ and $\bm{v}_2$ lie on $\bm{B}$, while $\bm{v}_1^{\prime}$ and $\bm{v}_2^{\prime}$ lie on $\bm{\sigma}_{12}$.}
        \label{fig:3dvector}
    \end{figure}

\begin{figure}
        \centering
            \includegraphics[width=1\linewidth]{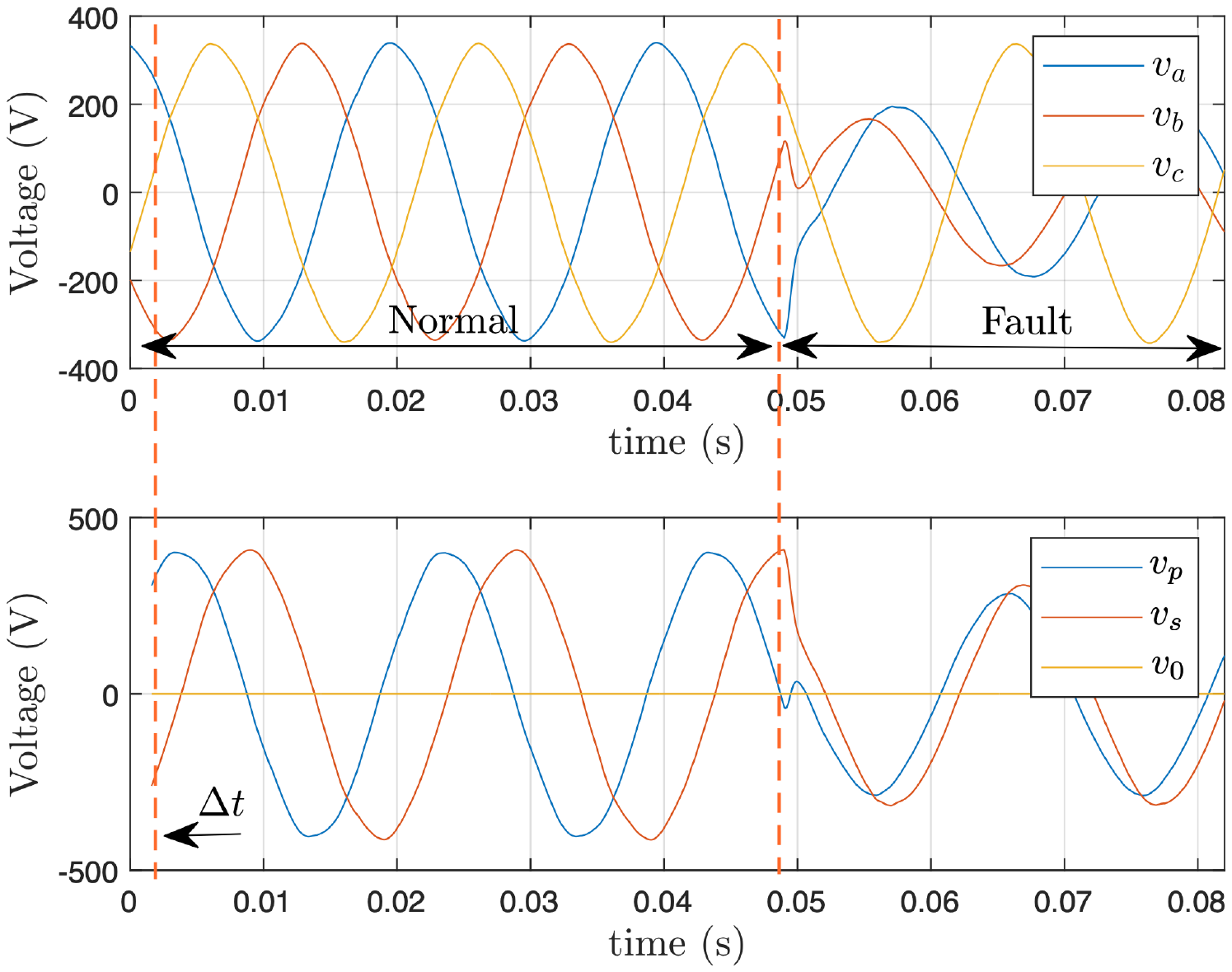}
        \caption{Experimental test under normal and fault conditions. Transformation of three-phases voltage system to a two-phase system.}
        \label{fig:resultsLab}
    \end{figure}

\begin{figure}
        \centering
            \includegraphics[width=1\linewidth]{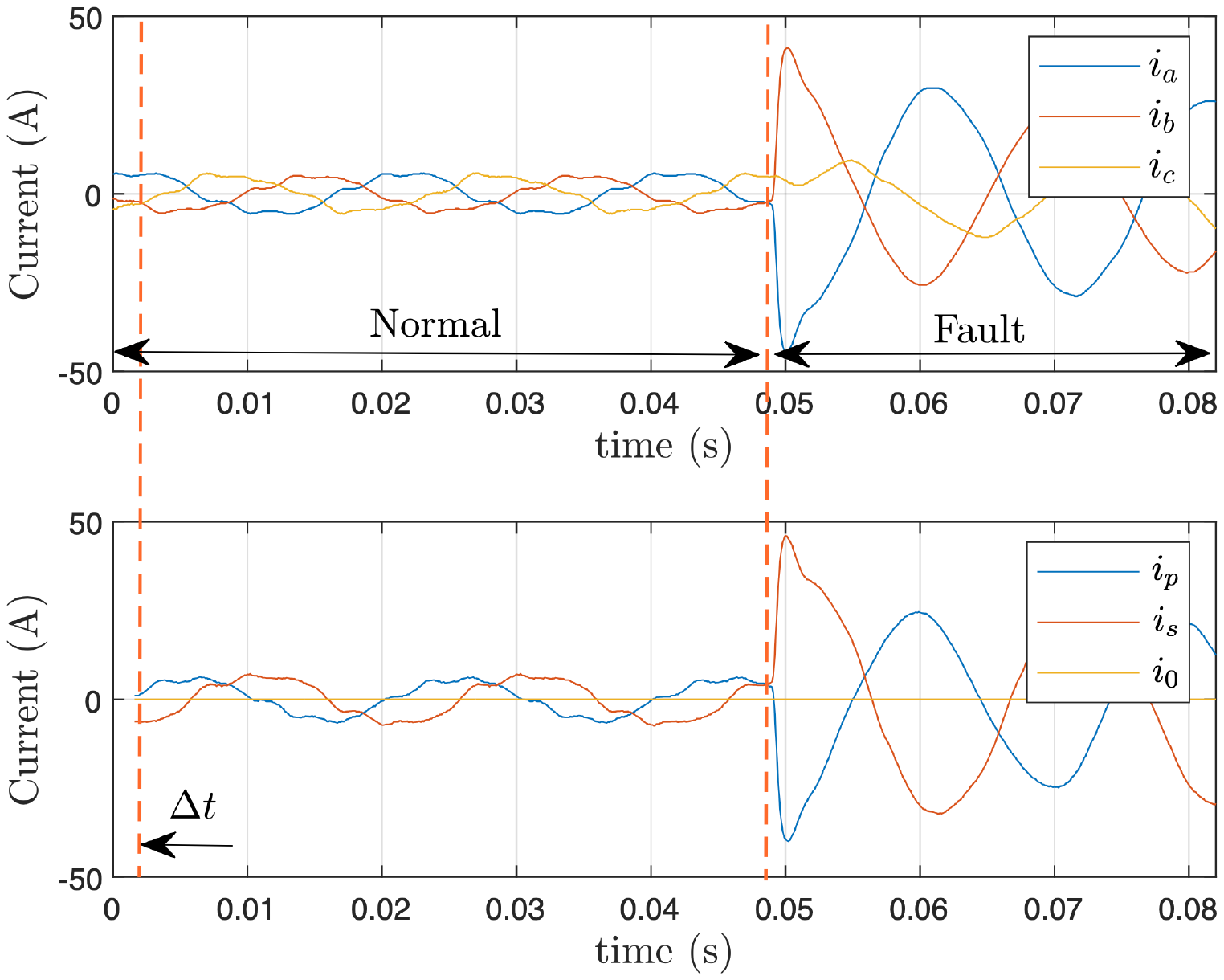}
        \caption{Experimental test under normal and fault conditions. Transformation of three-phases current system to a two-phase system.}
        \label{fig:resultsiLab}
    \end{figure}

    The voltage was measured with a sampling time of 100 $\mu$s on the source side. In this experiment, the voltages used to calculate the plane have a separation $\Delta t = 1.6$ ms. The voltage was taken at the initial time $t_1 = 0$ with the three-phase voltage vector  $\bm{v}_1 = \bm{v}(t_1) = 333.2031 \bm{\sigma}_1
 -198.0469\bm{\sigma}_2
  -135.1562\bm{\sigma}_3$ and with the second voltage vector at $t_2 = 1.6$ ms, with $\bm{v}_2 = \bm{v}(t_2) =270.3125\bm{\sigma}_1
 -297.2656\bm{\sigma}_2
  +26.9531\bm{\sigma}_3$.
  Hence, the angle of rotation $\theta$ and the rotor are presented in \eqref{eq:thetalab} and \eqref{eq:rotorlab}.
    \begin{align}\label{eq:thetalab}
        \theta &= 2.1863\mbox{\ rad.}\\
\label{eq:rotorlab}
        \bm{R} &= 0.4597 +0.6280\bm{\sigma}_{13} + 0.6280 \bm{\sigma}_{23} 
    \end{align}
    Fig. \ref{fig:3dvector} illustrates the geometry of this scenario. The vectors \(\bm{v}_1\) (red) and \(\bm{v}_2\) (black dashed) initially lie in the \(ps\) plane, represented by the bivector \(\bm{B}\). After a rotation of \(\theta\) radians around the intersection line between $\bm{B}$ and $\bm{\sigma}_{12}$, the \(ps\) plane aligns with \(\bm{\sigma}_{12}\), yielding the transformed vectors \(\bm{v}_1^{\prime}\) (blue) and \(\bm{v}_2^{\prime}\) (green dashed).
    The transformation results for the whole signal are shown in Fig. \ref{fig:resultsLab} and Fig. \ref{fig:resultsiLab}. 
    Figure \ref{fig:resultsLab} shows the three-phase voltages for the pre-fault i.e. normal and fault conditions. The voltage is used to compute the plane within 10$\%$ of a cycle, once the plane is computed, it is possible to calculate the angle and the rotor to transform the following samples. The lower part of Fig. \ref{fig:resultsLab} illustrates the three-phase voltages transformed into the $ps$ two-phase voltage system. The results confirm the robustness of the transformation method under experimental conditions. Fig. \ref{fig:resultsLab} also shows that the resulting voltages exhibit noticeable deviations from the ideal sinusoidal waveforms. Fig. \ref{fig:resultsiLab} presents the three-phase currents alongside their $ps$ transformed counterparts, demonstrating the method’s capability to handle typical experimental conditions, including harmonic distortion. Both voltage and current transformations showed computational efficiency and structural simplicity.


    \subsection{Extension to Six-Phase System}
    
    In this section, we present a comprehensive example of a six-phase system. This validation serves multiple purposes: it showcases the natural extension of our approach beyond three dimensions, demonstrates the numerical stability of the two-step rotation process, and provides detailed insight into the geometric transformations involved.

    \subsubsection{System Configuration and Initial Measurements}
    
    We consider a synthetic six-phase system where voltage measurements were taken at two distinct time instants, yielding the following voltage vectors in the six-dimensional space:

    \begin{alignat*}{6}
        \bm{v}_1 &= \phantom{1.00}\bm{\sigma}_1 &{}+{}& 1.7\bm{\sigma}_2 &{}-{}& 0.5\bm{\sigma}_3 &{}-{}& 0.5\bm{\sigma}_4 &{}+{}& 0.5\bm{\sigma}_5 &{}-{}& \bm{\sigma}_6 \\
        \bm{v}_2 &= 0.37\bm{\sigma}_1 &{}+{}& 0.7\bm{\sigma}_2 &{}+{}& 0.9\bm{\sigma}_3 &{}-{}& 0.1\bm{\sigma}_4 &{}-{}& 0.4\bm{\sigma}_5 &{}+{}& \bm{\sigma}_6
    \end{alignat*}

    These vectors represent synthetic unbalanced conditions where each phase exhibits different magnitudes and the spatial trajectory deviates significantly from any canonical plane.

    \subsubsection{Plane Identification and Bivector Analysis}
    
    Following our proposed methodology, the bivector $\bm{B}$ representing the voltage locus plane is computed using the outer product of the two measurement vectors:

    \begin{align*}
    \bm{B} &= \bm{v}_1 \wedge \bm{v}_2 =\nonumber \\ 
    &\phantom{-\,\,\,} 0.071\bm{\sigma}_{12} + 1.085\bm{\sigma}_{13} + 0.085\bm{\sigma}_{14} - 0.575\bm{\sigma}_{15} \nonumber \\
    &+ 1.370\bm{\sigma}_{16} + 1.880\bm{\sigma}_{23} + 0.180\bm{\sigma}_{24} - 1.013\bm{\sigma}_{25} \nonumber \\
    &+ 2.400\bm{\sigma}_{26} + 0.500\bm{\sigma}_{34} - 0.255\bm{\sigma}_{35} + 0.400\bm{\sigma}_{36} \nonumber \\
    &+ 0.245\bm{\sigma}_{45} - 0.600\bm{\sigma}_{46} + 0.110\bm{\sigma}_{56}
    \end{align*}
    
    The resulting bivector contains all fifteen possible basis bivectors in six-dimensional space, confirming that the voltage locus plane has a general orientation that does not align with any of the canonical coordinate planes. 

    \subsubsection{Two-Step Rotation Process Implementation}
    
    Since we are working in six dimensions where arbitrary planes may not intersect, we employ the two-step rotation methodology described in Section \ref{sec:method}.

    \paragraph{Step 1. Vector Alignment to $\bm{\sigma}_1$}    We first normalize $\bm{v}_1$ to obtain the unit vector
    \begin{align}
        \hat{\bm{v}}_1 &= 0.421\bm{\sigma}_1 + 0.716\bm{\sigma}_2 - 0.211\bm{\sigma}_3 - 0.211\bm{\sigma}_4 \nonumber \\
        &+ 0.211\bm{\sigma}_5 - 0.421\bm{\sigma}_6
    \end{align}
    
    The rotor $\bm{R}_1$ that aligns $\hat{\bm{v}}_1$ with $\bm{\sigma}_1$ is computed according to equation \eqref{eq:rotorR1}:
    \begin{align}
    \bm{R}_1 = &\phantom{+} 0.843 + 0.425\bm{\sigma}_{12} - 0.125\bm{\sigma}_{13} - 0.125\bm{\sigma}_{14} \nonumber \\
    &+ 0.125\bm{\sigma}_{15} - 0.250\bm{\sigma}_{16}
    \end{align}
    
    To verify the effectiveness of this rotation, we apply $\bm{R}_1$ to $\hat{\bm{v}}_1$
    \begin{equation}
    \bm{R}_1\hat{\bm{v}}_1\bm{R}_1^\dagger = \bm{\sigma}_1 + \varepsilon
    \end{equation}
    where $|\varepsilon|$ represents the numerical error with magnitude $2.22 \times 10^{-16}$, demonstrating the high numerical precision of the transformation.

    \paragraph{Step 2. Plane Alignment to $\bm{\sigma}_{12}$}  After transforming the original bivector using $\bm{R}_1$, we obtain an intermediate unit bivector $\hat{\bm{B}}_{\times}$
    \begin{equation}
    \hat{\bm{B}}_{\times} = \bm{R}_1 \hat{\bm{B}} \bm{R}_1^{\dagger}
    \end{equation}
    
    By construction, $\hat{\bm{B}}_{\times}$ now contains $\bm{\sigma}_1$ and intersects with the target plane $\bm{\sigma}_{12}$. The second rotor $\bm{R}_2$ is computed to align $\hat{\bm{B}}_{\times}$ with $\bm{\sigma}_{12}$ using \eqref{eq:rotorR2}:
    \begin{align}
    \bm{R}_2 = &\phantom{+} 0.813 + 0.363\bm{\sigma}_{23} - 0.018\bm{\sigma}_{24} \nonumber \\
    &- 0.169\bm{\sigma}_{25} + 0.421\bm{\sigma}_{26}
    \end{align}

    \subsubsection{Complete Transformation and Verification}
    
    The composite rotor that performs the complete transformation is:
    \begin{align}
    \bm{R} = \bm{R}_2\bm{R}_1 = &\phantom{+} 0.686 + 0.176\bm{\sigma}_{12} - 0.256\bm{\sigma}_{13} - 0.094\bm{\sigma}_{14} \nonumber \\
    &+ 0.173\bm{\sigma}_{15} - 0.382\bm{\sigma}_{16} + 0.306\bm{\sigma}_{23} \nonumber \\
    &- 0.015\bm{\sigma}_{24} - 0.142\bm{\sigma}_{25} + 0.355\bm{\sigma}_{26} \nonumber \\
    &- 0.048\bm{\sigma}_{1234} + 0.024\bm{\sigma}_{1235} - 0.038\bm{\sigma}_{1236} \nonumber\\
    &-0.023\bm{\sigma}_{1245} +0.057\bm{\sigma}_{1246} -0.010\bm{\sigma}_{1256}
    \end{align}

    To validate the complete transformation, we verify several key properties:
    
    \paragraph{Bivector alignment} The transformed bivector $\bm{B}' = \bm{R}\hat{\bm{B}}\bm{R}^{\dagger}$ yields
    \begin{equation}
    \bm{B}' = \bm{\sigma}_{12} + \varepsilon_B
    \end{equation}
    with numerical error $|\varepsilon_B| = 2.44 \times 10^{-16}$, confirming perfect alignment with the target plane.
    
    \paragraph{Vector transformations} The original measurement vectors transform as
    \begin{align}
    \bm{v}_1' &= \bm{R}\bm{v}_1\bm{R}^{\dagger} = 2.375\bm{\sigma}_1 \\
    \bm{v}_2' &= \bm{R}\bm{v}_2\bm{R}^{\dagger} = -0.015\bm{\sigma}_1 + 1.612\bm{\sigma}_2
    \end{align}
    
    The transformed vectors lie entirely within the $\bm{\sigma}_{12}$ plane, confirming successful dimensional reduction from six to two dimensions.

        \section{Conclusion}
		In this paper,  a novel geometric algebra framework to transform unbalanced multi-phase quantities has been introduced. Using bivector and geometric rotor elements, our approach directly identifies the $ps$-plane containing the vector space signal locus and transforms it to a canonical reference plane.
		
		The proposed method provides a robust generalization of the Clarke transformation that is valid for any degree of unbalance and requires only two voltage/current measurements. Unlike the classical Clarke transformation, which is tied to a specific plane with normal vector [1,1,1], our method works with the actual plane containing the voltage locus, ensuring that the transformed quantities lie entirely in a two-dimensional space.
		
		A significant advantage of our approach is its natural extensibility to $n$-dimensional spaces. While existing methods are intrinsically bound to three-phase systems, our geometric algebra framework can be directly applied to multi-phase systems with arbitrary number of phases. This extension is mathematically elegant and does not require fundamental reformulation of the theory, as the concepts of bivectors and rotors remain well-defined in $n$-dimensional spaces.
		
		The practical validation through power electronics converter control demonstrates the method's effectiveness in real control applications. The transformation eliminates oscillations in the zero component that occur with Clarke transformation under unbalanced conditions, while enabling the use of geometric power concepts through the geometric product of voltage and current vectors.
		
		The key advantages of our approach include its robustness against special cases, computational simplicity, clear geometric interpretation, and generality. These advantages make it well-suited for practical implementation in power electronic control systems, real-time monitoring and protection strategies, and general multi-phase systems with more than three phases.

\bibliographystyle{IEEEtran}
\bibliography{references.bib}

\end{document}